\documentclass[conference]{IEEEtran}
\IEEEoverridecommandlockouts
\usepackage{cite}
\usepackage{amsmath,amssymb,amsfonts}
\usepackage{enumitem}
\usepackage{booktabs}
\usepackage{algorithmic}
\usepackage{graphicx}
\usepackage{float}
\usepackage{textcomp}
\usepackage{xcolor}
\usepackage{epsfig}
\usepackage{bm}
\def\BibTeX{{\rm B\kern-.05em{\sc i\kern-.025em b}\kern-.08em
    T\kern-.1667em\lower.7ex\hbox{E}\kern-.125emX}}
\begin{document}

\title{Machine Learning-Driven Adaptive Power Allocation for Optical Wireless Networks\\
}
\author{\IEEEauthorblockN{ Walter Zibusiso Ncube, Ahmad Adnan Qidan, Taisir El-Gorashi and Jaafar M. H. Elmirghani}
\IEEEauthorblockA{\textit{Department of Engineering, Faculty of Natural, Mathematical and Engineering Sciences} \\
\textit{King’s College London, United Kingdom}\\
(walter.ncube, ahmad.qidan, taisir.elgorashi, jaafar.elmirghani)@kcl.ac.uk}
}

\maketitle
\begin{abstract}
Vertical-Cavity Surface-Emitting Lasers (VCSELs) have gained popularity in Optical Wireless Communication (OWC) due to their high modulation bandwidth, narrow spectral width, and directional beam, offering improved spectral efficiency and reduced multipath dispersion compared to Light-Emitting Diodes (LEDs). In this work, we explore the deployment of VCSELs as Access Points (APs) in an indoor environment under mobility and time-varying user distributions. To enhance performance, a Merged Access Point (MAP) topology is introduced to extend the serving area of each cell, whilst Zero Forcing (ZF) precoding is employed for inter-user interference management. A sum-rate maximisation problem is then formulated to maintain high-quality network operation in the dynamic environment. Although deterministic methods can solve the formulated problem, they become impractical in real time due to computational complexity, particularly under high user mobility and rapidly changing channel conditions. To address this, we propose a hybrid Machine Learning (ML)-based solution combining a low-complexity distance-based user association algorithm with a Convolutional Neural Network (CNN) for adaptive power allocation. Simulation results show that the proposed hybrid association–CNN framework achieves near-optimal performance while substantially reducing computation complexity relative to optimisation-based schemes. Furthermore, it operates in real time, with measured median and P95 inference latencies in the millisecond range, and maintains a small empirical worst-case gap to the Mixed Integer Linear Programming (MILP) optimum, demonstrating both practicality and robustness under mobility.

\end{abstract}


\section{Introduction}\label{sec1}

Optical Wireless Communication (OWC) has emerged as an enabler for next-generation networks, offering unregulated bandwidth for better performance. Early OWC systems using Light Emitting Diodes (LEDs) were constrained by bandwidth. However, advancements have introduced laser-based OWC systems, particularly Vertical-Cavity Surface-Emitting Lasers (VCSELs), which deliver focused beams for efficient power use, higher data rates, and longer transmission distances \cite{11164782}.
The narrow beam divergence of VCSEL transmitters enables high optical efficiency but results in limited cell coverage. Achieving full coverage therefore requires deploying a dense grid of VCSEL-based Access Points (APs), which in turn increases Inter-Cell Interference (ICI). To mitigate ICI and improve coverage continuity, neighbouring VCSEL APs can be merged to form larger logical cells and jointly serve users. This configuration is referred to as a Merged Access Point (MAP) topology \cite{7437435}. MAP helps smooth spatial coverage, reduce handovers under mobility, and enable distributed power transmission across multiple low-power APs rather than increasing the output of a single unit, an important consideration under optical eye-safety constraints. For context, similar coordination in Radio Frequency (RF) systems is achieved through techniques such as Coordinated Multipoint (CoMP) and cell-free MIMO. However, laser-based OWC differs fundamentally: transmit power is limited by eye-safety regulations, propagation is highly directional and Line-of-Sight (LoS)-dependent, and link quality is sensitive to receiver orientation, while diffuse reflections contribute minimally compared to RF multipath. However, despite the MAP topology bringing notable performance and robustness benefits, the larger cells demand efficient power and resource management to maintain a balance between performance and sustainability.

In general, efficient resource management is crucial for maximising network performance. Optimisation problems aimed at rate maximisation have been extensively investigated to  enhance the efficiency of OWC networks. For instance, resource allocation strategies were  explored in \cite{5547195,10072996,10621108} to optimise system performance. Specifically, in a power allocation optimisation problem was formulated in \cite{10621108} to maximise the aggregate data-rate of multi-stationary users in a VCSEL-based OWC network. Such optimisation problems  can be linearised and then  solved as a Mixed-Integer Linear Program (MILP). However, such methods are inherently complex and computationally intensive with increasing network size, posing significant challenges for real-time implementation.

Machine Learning (ML) has emerged as a promising solution for NP-hard optimisation problems in various contexts\cite{RL001,RL003,9566383,9569123,deep_learning_paper,9839259,10279546,s24072021,11003870,9374979,10585312}. Within this landscape, Reinforcement Learning (RL) has received considerable attention due to its ability to learn adaptive policies that optimise long-term performance in dynamic environments. In \cite{RL001}, a Q-learning and SARSA based RL algorithm was developed to maximise sum-rate in an Intelligent Reflective Surface (IRS)-aided indoor OWC network by dynamically allocating optical APs and mirror elements. In \cite{RL003}, a deep Q-learning-based resource-allocation scheme was developed for NOMA visible light communication systems, jointly optimising power distribution and LED tuning to enhance sum-rate and energy efficiency. Although RL methods can remove the need for labelled optima and adapt online \cite{9566383}, they typically come at the cost of longer and less predictable training, careful reward and constraint design and they require careful exploration control to prevent unstable handovers and Quality of Service (QoS) violations \cite{9569123}. These challenges motivate the exploration of supervised learning approaches, which learn deterministic mappings from data and can offer more stable, predictable behaviour. Building on this rationale, recent studies have demonstrated that supervised deep-learning models can deliver efficient and reliable optimisation; notably, in \cite{deep_learning_paper}, a deep learning algorithm was employed for resource allocation in massive MIMO systems, achieving high performance with minimal computational overhead. Similarly, in \cite{9839259}, an Artificial Neural Network (ANN) model was introduced in an OWC network to allocate fractional time resources based on traffic demands of stationary users with low computational overhead in real-time. Furthermore, in \cite{10279546} cooperative ANNs were designed for sum rate maximisation in a discrete-time laser-based OWC system serving stationary users. In \cite{s24072021}, Deep Neural Networks (DNN) were applied to handover and association decisions in hybrid LiFi/Wi-Fi, achieving robust decisions under mobility, while broader surveys, \cite{11003870}, document supervised ANN use for power/time allocation in OWC. Complementary work shows Convolutional Neural Networks (CNN) supporting OWC positioning/orientation estimation \cite{9374979,10585312}, evidence that learned models can exploit spatial structure and device orientation effects that matter for link quality and allocation. 
Compared with generic DNN architectures, CNNs exploit local spatial correlations, often achieving similar accuracy with fewer parameters and faster inference which is useful for real-time OWC \cite{11003870}. In contrast, Graph Neural Networks (GNN) can model user–cell graphs and support association/scheduling. However, they introduce graph construction overheads and higher complexity particularly on large-scale networks which limits their real-time deployment \cite{QGNN}. Despite these advancements, the application of ML in OWC networks remains an evolving field, particularly for dynamic scenarios involving user mobility in real-time. User mobility introduces rapid and unpredictable changes in channel conditions. Therefore, in multi-mobile user environments, the channel coherence time is relatively short, highlighting the need for ML models capable of making instantaneous decisions for performance enhancement \cite{coherence_time}.

In contrast to existing literature, this work models multi-mobile users in an OWC network, where users move between randomly selected waypoints at a given speed, creating continuously varying link distances and user--cell associations. In such environments, optimisation-based methods used in \cite{5547195,10072996,10621108} become impractical due to the dynamic nature of the network. Although trained ML models can provide fast online decisions, their effectiveness depends on training data that accurately capture the dynamic behaviour of mobile users, and therefore, generating  large and full datasets for training poses a scalability challenge. To address these challenges while avoiding full optimisation during operation, a hybrid ML-based scheme is adopted in which user association is handled by a low-complexity distance-threshold heuristic, and power allocation is learned using a CNN trained offline on  mixed-integer linear programming (MILP)-optimal power samples. This decoupling reduces the dimensionality of the learning problem, enables efficient dataset generation, and provides a scalable, mobility-aware solution suitable for real-time operation in dynamic OWC environments. 
The choice in the hybrid scheme is motivated by the fact that given our indoor system and spatially structured inputs, a CNN-based hybrid offers a balanced trade-off among accuracy, latency, and implementation simplicity for user connectivity and resource management, while other ML variants remain promising for future extensions that jointly consider user association and power allocation. Therefore, the main contributions of this work are listed as follows:

\begin{itemize}
\item A sum-rate maximisation problem is formulated for MAP-based multi-mobile user laser OWC networks, where user–cell association and per-cell power allocation are jointly optimised. The resulting problem is NP-hard, and a direct solution becomes computationally infeasible for dynamic scenarios. To address this, the original optimisation problem is decomposed into a user association subproblem and a power allocation subproblem, which are solved separately with significantly reduced complexity.
\item A low-complexity distance-threshold association algorithm is proposed to generate deterministic user–cell associations. This decomposition reduces the dimensionality of the continuous optimisation, stabilises association under mobility, and produces structured inputs that can be directly fed into a learning-based power allocation model.
\item A CNN-based power allocation model, trained offline on MILP-derived optimal power allocations, is introduced to compute the continuous power vector using the association output as input. The CNN leverages the spatial layout of MAP cells and their aggregated feature representations to approximate the optimal solution without requiring MILP solving, with low computational complexity and high accuracy.
\item The hybrid ML-based scheme provides a scalable control solution that delivers near-optimal sum-rate performance, achieves millisecond-level inference latency, and exhibits a small empirical worst-case gap relative to the MILP optimum. These results demonstrate the practicality and robustness of the approach for dynamic indoor OWC deployments.

\end{itemize}

It should be noted that unlike our earlier MILP-based study \cite{10621108}, which focused on stationary users and did not consider deployment feasibility, the present work extends the analysis to a VCSEL-based MAP architecture under user mobility and integrates a CNN-driven power allocation scheme to enable low-complexity decision-making in dynamic environments. Rather than targeting exact optimality in static scenarios, this work emphasises practical applicability by combining heuristic user association with learning-based power control in a mobility-aware OWC system.

\section{System Model}

We consider an indoor laser-based OWC network designed to deliver high data rate communication. $\mathcal{A},  a = \{1, \dots, A\}$,  optical APs are located on the ceiling where each optical AP is designed as a micro lens VCSEL array for data transmission. The APs are arranged into $\mathcal{C}, c = \{1, \dots, C\}$, multiple optical cells and serve $\mathcal{U}, u = \{1, \dots, U\}$, users, pointing upward and located on the communication plane, with time-varying  distribution and traffic demands. We define the set of users within the coverage area of cell $c$ as $\mathcal{U}_c, u = \{1, \dots, U_c\}$. Users are equipped with an Angle Diversity Receiver (ADR) comprising multiple receiving faces, where each face uses a Compound Parabolic Concentrator (CPC) coupled to a small photodiode (PD) array to enhance optical concentration and Field of View (FoV). To maintain high receiver bandwidth, each PD is followed by an individual Transimpedance Amplifier (TIA), and the photocurrents within each face are combined using Equal-Gain Combining (EGC), providing spatial diversity without compromising bandwidth as in \cite{10299714}. The optical wireless channel is modelled considering only the dominant LoS component, assuming that each receiver is oriented toward its serving transmitter and that the optical path remains unobstructed.\footnote{In practice, optical channels may experience additional effects such as receiver orientation variability, temporary LoS blockage, and surface reflections. These factors can introduce attenuation or short-term link degradation but are not explicitly modelled here, as their inclusion would require a dynamic orientation and blockage model. Diffuse reflections are considered negligible due to the high directionality of VCSEL beams. Incorporating these non-LoS and reflection effects is reserved for future extensions of this work.} Furthermore, the system assumes perfect Channel State Information (CSI)\footnote{In line with the majority of OWC resource allocation studies, we assume near-perfect CSI at the transmitter. In practice, CSI may be impaired by estimation errors or feedback delay; however, analysing the impact of imperfect CSI on the proposed framework is beyond the scope of this work and is left for future investigation.} is available at the transmitter for power allocation and Zero Forcing (ZF) precoding.
This simplifies the analysis and allows us to benchmark ideal performance. All the optical APs, together with a WiFi AP for uplink transmission, are connected to a Central Unit (CU), which provides AP-coordination and resource management.
\subsection{User mobility and cell configurations }

We model user mobility using the Random Waypoint model where each user moves between randomly selected waypoints with speed $\nu$  as in \cite{RWP_Paper}. Moreover, the arrival of user requests is represented by a Poisson process with arrival rate $\mu$, and the duration of these demands follows a hyper-exponential distribution to account for variability in user behaviour as in \cite{PDP_paper_2}.
\begin{figure}[h]
\centering
\includegraphics[width=1\textwidth, height=0.33\textheight, keepaspectratio]{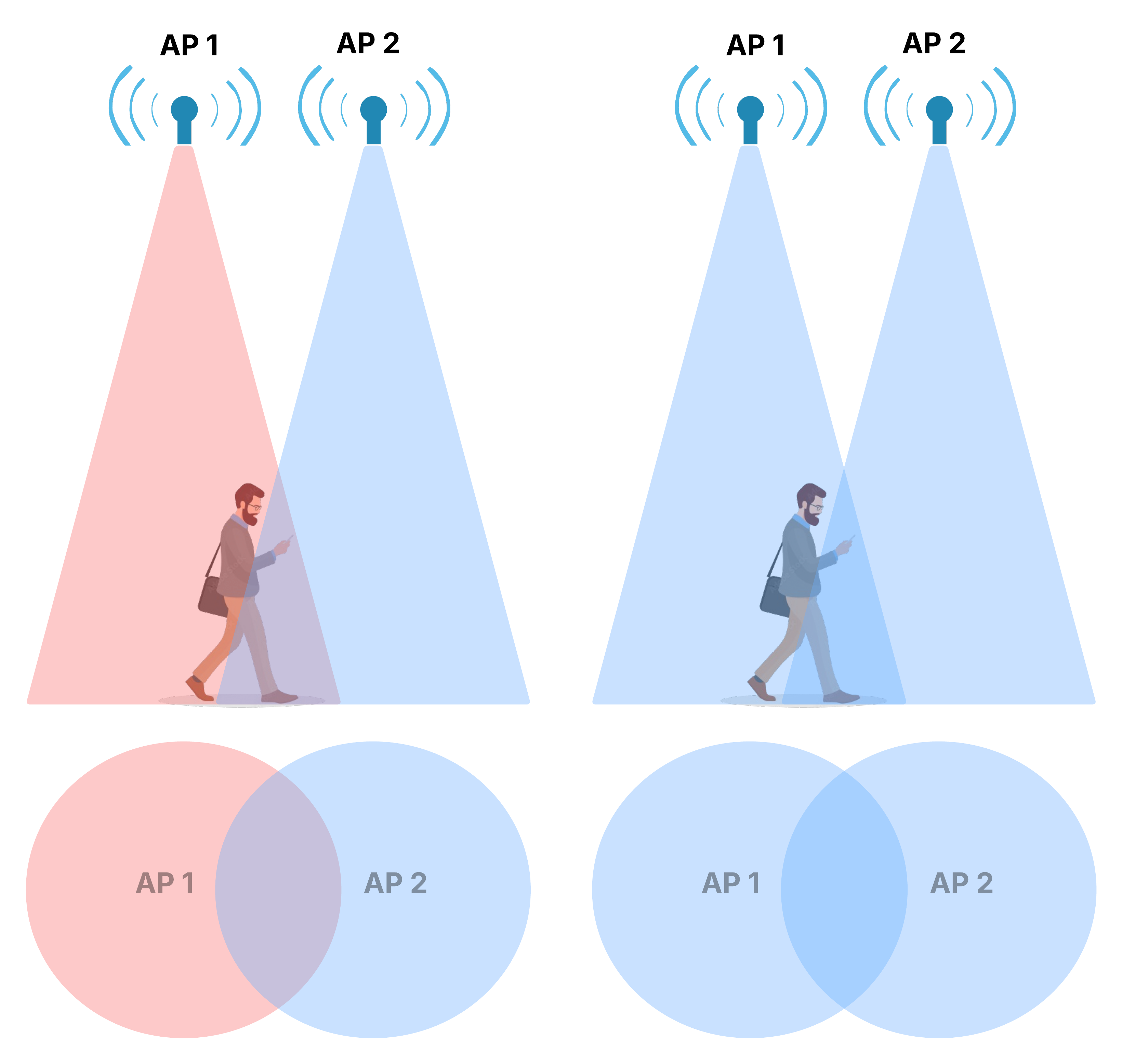}
\caption{ Traditional  \textit{(left)} and  MAP \textit{(right)} cell formations.}
\vspace{-15pt}
\label{MAP Topologies}
\end{figure}
 The system supports two distinct cell configurations: the Traditional cell formation and the Merged Access Point cell formation. In the traditional system, each AP independently defines a discrete optical cell. While this design is simple and effective for small setups, it has challenges in larger networks due to ICI from overlapping LoS beams as shown in Fig. \ref{MAP Topologies} (left). Note that, narrowing the receiver’s FoV can reduce ICI but leads to coverage gaps and frequent handovers, causing performance degradation, i.e., low QoS \cite{7249136}. To address this, we introduce the MAP configuration as in our previous work \cite{10621108}, which combines adjacent APs into a larger optical cell as depicted in Fig. \ref{MAP Topologies} (right). The system uses ZF for interference management, ensuring seamless coverage and improved user experience \cite{10621108}, \cite{7249136}. For more information, we refer readers to \cite{10621108}.

\subsection{Transmit optical power}

We assume the VCSEL beam exhibits a Gaussian profile which is characterised by two key parameters: the beam waist (\( w \)) and the wavelength (\( \lambda \)). For a Gaussian beam propagating along the \( z \)-axis, the intensity distribution is given by:

\begin{equation}
I(r, z) = \frac{2P_t}{\pi w^2(z)} \exp\left(-\frac{2r^2}{w^2(z)}\right),
\end{equation}

\noindent where \( P_t \) is the transmitted optical power, \( r \) and \( z \) represent the radial and axial positions, respectively, and \( w(z) \) is the beam spot radius as a function of \( z \). In this study, we consider the parameters of the Gaussian beam after being refracted through a lens as in \cite{10621108}. The total transmitted power from an \( n \times n \) VCSEL array is the cumulative output power of all individual VCSELs of the array, $P^a = \sum_{i=1}^{n \times n} P_t^i$, where \( P_t^i \) represents the transmitted power of the \( i \)-th VCSEL. To ensure compliance with eye safety standards, the total emitted power from an AP or array must not exceed the permissible limits, \( P_{\text{safe}} \), such that $P^a\leq P_{\text{safe}}$ \cite{10621108}.

\section{Sum-rate Optimisation Problem}
We consider ZF to mitigate multi-user interference (MUI) effectively within the coverage area of each optical cell $c$. Thus, the received signal can be expressed as:

\begin{equation}
\textbf{y}^{c}(t) =  \textbf{H}^{c}    \textbf{g} (t) +\textbf{ I} (t) + \boldsymbol{v}(t),
\label{eq9}
\end{equation}

\noindent where \(\textbf{H} \in (\mathbb{R})^{U_c}\) is the channel gain matrix between the \(U_c\) users and the optical cell $c$. Moreover, \(\textbf{ I} (t)\) is the interference imposed on the users, and \(\boldsymbol{v}(t)\) is the receiver noise vector.  The transmitted signal is defined  as; $\textbf{g}(t) = \textbf{W} \cdot \textbf{X} (t)$, where \(\textbf{X}(t) \in (\mathbb{C})^{U_c}\) is the multi-user symbol vector, and the ZF precoding matrix \(\textbf{W} \in (\mathbb{C})^{U_c}\) is explicitly given by the pseudo-inverse of the channel matrix \(\textbf{H}^c\), i.e., $\textbf{W} = \textbf{H}^+$, where \(\textbf{H}^+\) is the pseudo-inverse of \(\textbf{H}^c\). From (\ref{eq9}), 
 the received signal of specific user \(u \in \mathcal{U}_c\) after removing the DC bias imposed to ensure signal non-negativity is given by:
 
\begin{equation}
y^{c}_{u} (t)= R_{\text{PD}} H_{u}^c  \sqrt{P^{c}_{u}} x_{u}(t)+ I_{u} (t) + v_{u} (t),
\label{eq10}
\end{equation}

\noindent where $R_\text{PD}$ is the responsivity, and $H_{u}^c= \sum^{PD}_{pd=1} \sum^{A_c}_{i=1}H_{ipd}^u$, where $A_c$ is the number of optical APs forming cell $c$. Moreover, $P_{\text{u}}^{c}$ is the power allocated to user $u \in \mathcal{U}$ to transmit $x_{u}(t)$, $I_{u} (t)$ is the interference imposed on user $u$, and \(v_{u} (t)\) is the noise on user $u$. The receiver noise is modelled as additive white Gaussian noise with variance 
$v_u^2 = N_0 B$ \cite{10299714}, where $B$ is the electrical bandwidth and $N_0$ is the single-sided
noise power spectral density. The total noise PSD accounts for thermal noise, shot noise,
and VCSEL Relative Intensity Noise (RIN), and is given by: 
\begin{equation}
    N_0 \;=\; 
\frac{4 k T F_n}{R_F}
\;+\;
2 q_e R_{\mathrm{PD}} P^{c}_{u}
\;+\;
\mathrm{RIN}\, (R_{\mathrm{PD}} P^{c}_{u})^2 ,
\end{equation}

\noindent where $k$ is Boltzmann’s constant, $T$ is the receiver temperature, $R_F$ is the TIA 
feedback resistance, $F_n$ is the TIA noise figure, $q_e$ is the electron charge, 
$R_{\mathrm{PD}}$ is the photodiode responsivity, and $P_{\mathrm{rx}}$ is the received optical power.

At this point, the user $u$ served by cell $c \in \mathcal{C}$ receives the data rate $R_u^c$, which can be determined using the lower bound of the Shannon capacity formula \cite{Data_Rate_Paper,qidan2025},

\begin{equation}
\begin{split}
R_u^c
= B \,\log_{2}\Biggl( 
1 \;+\; \frac{\exp(1)}{2\pi} \cdot \,\frac{ (R_{\text{PD}}H_{u}^c)^{2}\,P_{u}^{c}}
               {\Bigl(\,\,v_{u}^{2} 
                  \;+\; \sum_{\substack{c' \in \mathcal{C} \\ c' \neq c}} P_{u}^{c'}\Bigr)}
\Biggr).
\end{split}
\label{eq15}
\end{equation}

We now formulate an optimisation problem based on maximising the system sum rate as follows:

\begin{equation}
\max \sum_{u \in \mathcal{U}} \sum_{c \in \mathcal{C}}\log \left(  S_u^c R_u^c \right)
\label{eq17}
\end{equation}

subject to:
\vspace{-10pt}

\begin{equation}
\sum\nolimits_{c=1}^C S_u^c = 1, \quad \forall u \in \mathcal{U},
\label{eq25}
\end{equation}

\begin{equation}
P_{u}^{\text{min}} \leq \sum\nolimits_{c=1}^C S_u^c P_u^c  \leq P_{u}^{\text{max}}, \forall u \in \mathcal{U},
\label{eq26}
\end{equation}

\begin{equation}
\sum\nolimits_{u=1}^{U_c} P_u^c \leq P^c \quad \forall c \in \mathcal{C},
\label{eq27}
\end{equation}
where \(S_u^c\) is a binary assignment term that denotes whether or not  user \(u\) is assigned to cell \(c\). Constraint (\ref{eq25}) ensures that each user is assigned to exactly one cell. (\ref{eq26}) ensures that the minimum demand for each user is met and if the cell is not overloaded, more power is allocated to the user until it reaches the maximum power permissible for the user. (\ref{eq27}) is a capacity constraint that ensures that the total power assigned to all the users $U_c$ within the coverage area of the cell does not exceed the cell's capacity. Note that, $P^c=P^a$ for the traditional cell configuration, and $ P^c= \sum_{a  \in \mathcal{A}_c  } P^a$ for the MAP cell configuration where $\mathcal{A}_c$ contains the APs belonging to cell $c$. The optimisation problem in (\ref{eq17}) is NP-hard  due to the coupling of \(S_u^c\) and $P_u^c$, and it has complexity of $\mathcal{O}(C^{U})$. Therefore, solving it directly in real time is infeasible, especially in dynamic OWC networks where user mobility constantly alters the channel conditions and association distributions. Learning models such as ANNs require large and diverse datasets to generalise well in real-time. In our network, a high-quality dataset must capture both user association and power allocation across a wide range of network states, which poses a significant challenge. In other words,  insufficient coverage of realistic network conditions or limited dataset size may lead to over fitting, where ANN models fail to generalise to unseen scenarios during deployment.
Therefore, we decompose it into two sub-problems:  user association and power allocation. The user association sub-problem is solved subject to constraint (\ref{eq25}) using predefined distance threshold-based search or heuristic algorithm, as follows:
\begin{enumerate}
\item Each  user, $u \in \mathcal{U} $, is associated with the closest optical cell, $c \in \mathcal{C}_{th}$, i.e.,
\begin{equation}
c = \arg\min\limits_{c \in \mathcal{C}_{th}} \mathrm{d}(u, c),
\end{equation}
where $\mathcal{C}_{th}$ contains $C_{th}$ optical cells, $C_{th} < C$, located at  a maximum distance less than or equal to a threshold distance $d_{th}$. Moreover,  $\mathrm{d}(u, c) \leq d_{th} $ denotes the Euclidean distance between user $u$ and optical cell $c$, where  $\mathrm{d}(u,c)= \sqrt{(\check{x}_{u}-\check{x}_{\zeta_{c}})^{2}+(\check{y}_{u}-\check{y}_{\zeta_{c}})^{2}}$, where $ (\check{x}_{u},\check{y}_{u}) $ are the coordinates of user $u$, and   $ \left(\check{x}_{\zeta_{c}}, \check{y}_{\zeta_{c}} \right) $ are the  centroid coordinates of cell $c$ denoted as $\zeta_{c}$. Note that, for the traditional cell formation, the centroid $\zeta_{c}$ is given by the coordinates of single optical AP, $ \left(\check{x}_{\zeta_{c}}, \check{y}_{\zeta_{c}} \right)=  (\check{x}_{a},\check{y}_{a}) $, where $(\check{x}_{a},\check{y}_{a}) $ are the coordinates of optical AP $a \in \mathcal{A}$. On the other hand, for MAP cell formation, the centroid $\zeta_{c}$ is determined by averaging the coordinates of the optical APs forming cell $c$, e.g., 

\begin{equation}
\label{eq:centr} 
\zeta_{c}=\left(\dfrac{\check{x}_{a}+\check{x}_{a^{'}}}{2} ,\dfrac{\check{y}_{a}+\check{y}_{a^{'}}}{2}\right),
\end{equation}
where cell $c$ is composed of two optical APs, $c=\{a,a'\}$, and $(\check{x}_{a'},\check{y}_{a'}) $ are the coordinates of optical AP $a' \in \mathcal{A}$. 

\item Once the users  associated to each cell, $\mathcal{U}_c$, $c \in \mathcal{C}$, are determined, the algorithm considers assigning other users that are not associated to any optical cell, and can be served  by optical cell $ c $. Thus, the elements of $\mathcal{U}_c$ is updated as 
 
\begin{equation}
\mathcal{U}_c=  \mathcal{U}_c \cup \{u \in \mathcal{U},  \mathrm{d}(\zeta_{c},u) \leq d_{th}\}.
\end{equation} 

\item The proposed algorithm ends  with assigning $ U $ users to $ C $ optical cells, while each optical cell serves a unique set of users, i.e., 

\begin{equation}
\label{uniq}
~\begin{matrix}
\mathcal{U}_{c} \cap \mathcal{U}_{c^{'}}=\emptyset, ( c \neq c^{'}), 
\big\{{c,c^{'}}\big\} \in C, \big\{\mathcal{U}_{c}, \mathcal{U}_{c^{'}} \big\} \in \mathcal{U}.
\end{matrix}
\end{equation}
 \end{enumerate}
The proposed algorithm has a very low complexity of $\mathcal{O}(U \times C_{th}), C_{th} < C $, which makes it effective  for instantaneous solutions in fast-changing environments.

The user association vector determined above is then used as  input into the power allocation sub-problem which is solved using an ANN trained to optimise the power allocation vector, as in the next section. We solve the two sub-problems iteratively, and in each iteration, the user association vector and the power allocation vector are alternately updated. This process continues until convergence is achieved, and a sub-optimal solution for the original optimisation problem is obtained. 

\begin{table}[h]
\centering
\scriptsize
\caption{System Parameters}\label{tab:system-parameters}

\resizebox{\linewidth}{!}{%
\begin{tabular}{@{}llll@{}}
\toprule
\textbf{Description} & \textbf{Value} & \textbf{Description} & \textbf{Value} \\
\midrule
Beam waist radius & $5\,\mu\mathrm{m}$ & VCSEL wavelength & 1550 nm \\
Laser RIN density & $-155\,\mathrm{dB/Hz}$ & Lens refractive index & 1.55 \\
Photodetector type & InGaAs PIN & Photodetector responsivity & $0.9\,\mathrm{A/W}$ \\
TIA transimpedance gain & $1.0$\;k$\Omega$ & TIA voltage-noise density & $1.5$\,nV/$\sqrt{\mathrm{Hz}}$ \\
TIA input-referred noise density & $5\,\mathrm{pA}/\sqrt{\mathrm{Hz}}$ & Feedback resistor & $1.0$\;k$\Omega$ \\
Receiver half-angle FoV & $30^\circ$ & Power per beam & 50 mW \\
TIA electrical bandwidth & 3.5 GHz & Validation dataset size & $10\%\ \text{of}\ N$ \\
ANN hidden layers & 3 & Training dataset size & $90\%\ \text{of}\ N$ \\
\bottomrule
\end{tabular}%
}
\end{table}

\section{Power Allocation Artificial Neural Network}
 We use a CNN, which is one of the common ANNs known for its efficiency and speed. In this work, the input layer of the CNN  corresponds to information from the system such as the user association vector and QoS demands. Moreover, the hidden layer has $B$ fully connected hidden layers and each layer has a total of $AN$ artificial neurons. We determine the output of each neuron, $z_{a,b}$, as follows \cite{9839259}:

\begin{equation}
z_{a,b} = \alpha_{a,b}\left[\textbf{w}_{a,b}\otimes z_{o,b-1} + \zeta_{a,b}\right],
\label{eq:neuron_output}
\end{equation}
where $\alpha[.]$ is a Rectified Linear Unit (ReLU) activation function,  $\textbf{w}_{a,b}$ is the weight vector of the $a$-th neuron in the $b$-th layer, $\otimes$ is the convolution operator of the CNN, $z_{o,b-1}$ is the output of layer $b-1$ and $\zeta_{a,b}$ is a scalar bias. Finally, the output layer determines the power allocated to each user and the  available power budget of each optical cell.  

We choose a CNN because the user--cell topology and association can be encoded as a spatial tensor where nearby entries are strongly correlated. Convolutions exploit this locality with shared kernels, achieving comparable accuracy to dense networks with significantly fewer parameters and lower inference latency. In contrast, fully connected DNNs disregard spatial structure and require more parameters to reach similar performance, while GNN/RL solutions introduce higher model and implementation complexity and, for RL, longer and less predictable training before stable policies emerge. While CNNs serve our objectives effectively, we acknowledge that alternative models such as GNNs, which can model graph-based user-cell relationships, Quantum Graph Neural Network (QGNN) that leverage the unique properties of quantum systems to reduce computational overhead \cite{QGNN, QGNN3}, or RL approaches for sequential decision-making \cite{RL001} are promising directions for future work.

It is also important to highlight that ML was not employed for user association, as constructing a reliable and robust training set for both user association and power allocation is difficult at our scales and timescales. In supervised learning, each training data would require a joint optimal label (association vector + power vector) under instantaneous constraints and mobility. Generating those labels means repeatedly solving a coupled, NP-hard problem across a wide state space of user positions, loads and AP configurations which can result in an exponential increase that is significantly more costly than deploying the heuristic introduced in Section 3. Moreover, association labels are often non-unique as there many near-equivalent assignments, which introduces label noise and affects convergence, while mobility induces distribution shift that would demand orders of magnitude more data to remain robust. Reinforcement learning can be used to avoid labels however this would require online exploration and careful reward shaping to prevent oscillatory handovers. In contrast, the proposed deterministic distance-threshold association is data-free, stable, and fast, and it reduces the learning problem’s dimensionality, allowing the CNN to focus on the continuous power allocation sub-problem where ML yields the acceptable runtime and performance.

\subsection{Dataset Generation}
We aim to train the ANN so that it establishes a mapping function $f(M;.)$ that allocates power based on user association, user demands, and other system constraints as in the optimisation problem. Let's assume $M$ as a set of weight terms that serve as a bridge between the input and output layers. In this context, we must generate a dataset for training the ANN to select the optimal set of parameters, \(M^*\), that provides the maximisation of the system’s sum rate in real-time scenarios.

We generate a dataset comprising \(N\) data points. Each data point \(n\) corresponds to the user association and user requirements within the range \(P^{\text{min}}_u \leq P_u \leq P^{\text{max}}_u\) to \(C\) cells. The power allocated to a specific user varies for each user based on their serving cells,  and their activity at a given time, i.e., \(P_u \neq P_{u'}\) for \(u \neq u'\). The dataset must also account for maximising the sum rate of the users. Therefore, following the determination of user association,  each cell \(c\) independently maximises the sum rate of its own users $U_c$ as follows
\begin{multline}
\max_{P_u^c} \Bigg[\sum_{u \in U_c} \log\left( R_u^c\right) - \omega_c \bigg(\sum_{u \in U_c} P_u^c -P^c\bigg) \\-\sum_{u \in U_c}\beta_{u,1} \bigg(P_u^c -P_{u}^{\text{max}}\bigg) + \sum_{u \in U_c}\beta_{u,1} \bigg(P_{u}^{\text{min}}-P_u^c\bigg) \Bigg],
\label{eq:utility_function}
\end{multline}
where \(\omega_c\), \(\beta_{1}\), and \(\beta_{2}\) are multipliers corresponding to the cell capacity and user requirement constraints. To solve this problem, the optimal power allocated to each user by a cell is determined using the Karush-Kuhn-Tucker (KKT) conditions \cite{KKT_book}. An iterative process is then applied to update the multipliers using gradient-based algorithms as follows:
\begin{equation}
\begin{aligned}
\omega_c^{(i+1)} &= \left[\omega_c^{(i)} - \varphi_\omega^{(i)}( \sum\nolimits_{u \in U} P_u^{c*}- P^c)\right]^+, \\
\beta_{u,1}^{(i+1)} &= \left[\beta_{u,1}^{(i)} - \varphi_{\beta_1}^{(i)} \left(P_u^{c*}-P^{\text{max}}_u \right)\right]^+, \\
\beta_{u,2}^{(i+1)} &= \left[\beta_{u,2}^{(i)} - \varphi_{\beta_2}^{(i)} \left( P^{\text{min}}_u- P_u^{c*}\right)\right]^+,
\end{aligned}
\label{eq:multiplier_updates}
\end{equation}
where \([.]^+\) represents a projection onto the positive orthant to ensure \(\omega_c, \beta_{u,1}, \beta_{u,2} \geq 0\), \(\varphi_j^{(i)}\) (\(j \in \{\omega_c, \beta_1, \beta_2\}\)) is the step size at iteration \(i\), \(P_u^{c*}\) is the optimal power allocation for user \(u\) from cell \(c\). After solving the optimisation problem for various user distributions and activities,  the maximised rates are recorded to create the ANN training dataset. It is worth mentioning that (\ref{eq:utility_function}) can be solved directly in real time under static conditions, however this becomes impractical in highly dynamic environments. When multiple users are moving in random directions at varying speeds, the rapid changes in network state make traditional optimisation too slow. In such cases, ML models trained to take instantaneous actions become essential for maintaining real-time performance.

\subsection{ANN Implementation}

The proposed network consists of two convolutional layers with 32 and 64 filters (kernel size $3\times3$), each followed by a ReLU activation and a $2\times2$ max pooling layer. A flattening layer and two fully connected layers with 128 and 64 neurons precede a linear output layer that predicts the power allocation vector. The ANN input is a fixed-size tensor that encodes, for each optical cell, the user association, the aggregated user QoS demand, and the available power budget. All continuous input features are normalised to the range $[0,1]$ to ensure that all inputs have a similar scale, which stabilises training and prevents some features from dominating others. Training is performed offline using supervised learning, where optimal power allocation vectors obtained from MILP solutions serve as ground-truth labels. The model is trained using the Adam optimiser with a learning rate of $10^{-3}$, mean-square-error loss, and a batch size of 64. Dropout and early stopping are applied to mitigate overfitting. Once trained, the CNN performs online inference through a single feed-forward pass to provide real-time power allocation decisions with minimal latency.

\subsubsection{Offline Phase}
In training of the CNN model, our aim is to identify the optimal set of weight terms, $M^*$, that aligns the CNN model's input and output. The training dataset consists of \(N\) data points, where the \(n\)-th data point is represented by the input \( \mathbf{v}(n)\) and the corresponding optimal power allocation \(\mathbf{P}^*(n)\). The output layer of the CNN estimates the power allocation vector $\hat{\mathbf{P}}(n)$.
 The CNN model learns by minimising a loss function between the estimated power allocation and the optimal allocation, i.e.,  $\min \frac{1}{N} \sum_{n=1}^N \Theta(\hat{\mathbf{P}}(n), \mathbf{P}^*(n))$, where \(\Theta(.,.)\) denotes the mean-square-error (MSE) function. In this way the CNN can be trained to provide sub-optimal solutions in time-varying scenarios, including those not explicitly covered in the training dataset.

\subsubsection{Real-Time Phase}
After training, the CNN is deployed to perform power allocation in real time. In this phase, the model determines the power allocation for users based on the determined user-cell association, user traffic demands, and available resources of the cells. Cells and users iteratively update their multipliers, $\omega_c, \beta_{u,1}$ and $\beta_{u,2}$, based on the output of the CNN model. These updates follow gradient-based methods to ensure that power allocation satisfies the constraints and maximises efficiency. Additionally, the CNN dynamically updates power allocation everytime user requirements and distributions change.

\section{Computational Complexity}
The joint MILP formulation jointly optimises binary user--cell associations and continuous power allocations, leading to a combinatorial search space of size $C^{U}$. Solving this problem using branch-and-bound incurs exponential-time complexity $O(C^{U})$ in the number of users and cells, where each node in the branch-and-bound search requires solving a continuous linear or quadratic subproblem, whose computational cost scales with the number of continuous variables. Consequently, the overall computational cost grows rapidly with network size, making MILP-based optimisation impractical for real-time deployment.

In contrast, the proposed hybrid ML-based scheme decomposes the problem into a distance-threshold association stage and a CNN-based power allocation stage. The association stage examines only $C_{\text{th}}\!\le\!C$ candidate cells per user, resulting in $O(U\,C_{\text{th}})$ complexity per iteration which scales linearly with the number of users.

The subsequent CNN inference for power allocation requires a single feed-forward pass, whose cost depends only on the network architecture and not on the instantaneous network state. For an input tensor of size $M$, a convolutional layer with kernel size $k$, input channel depth $C_{\text{in}}$, and output channel depth $C_{\text{out}}$ incurs $O(M\,k^{2}\,C_{\text{in}}C_{\text{out}})$ operations. 
With the compact two-convolution architecture used in this work, the overall CNN inference complexity grows approximately linearly with $M$ and remains constant for a trained model during deployment. It should be noted that in our architecture, the convolutional term dominates the total inference cost, and the contribution of the dense layers is negligible in comparison, so it is absorbed into the convolutional complexity term.

Therefore, the total complexity of the proposed framework can be expressed as
\begin{equation}
O(U\,C_{\text{th}}) + O(M\,k^{2}\,C_{\text{in}}C_{\text{out}}),
\end{equation}
where only the first term scales with the number of users. Thus, the proposed approach grows linearly with $\mathrm{U}$ and maintains effectively constant inference cost for the CNN.

\section{Results and Discussion}
We consider a $5\,\text{m} \times 5\,\text{m} \times 3\,\text{m}$ indoor environment with $A = 16$ APs uniformly distributed across the ceiling to serve $U = 8$ users  on the communication plane at a distance of $3\,\text{m}$ from the ceiling. This configuration is selected to balance realistic modelling with computational feasibility and allows benchmarking against optimisation-based baselines such as MILP without excessive runtime. Although the proposed hybrid ML-based scheme can be extended to larger networks, such scaling introduces non-negligible costs. Specifically, generating labelled training data becomes more expensive as each sample requires solving a higher-dimensional optimisation problem. Additionally, the computation of the zero-forcing within each MAP cluster scales as the number of users within the cluster due to the matrix pseudo-inverse operation. Larger spatial layouts would also require deeper or wider neural architectures, increasing memory usage, training time, and the risk of over-fitting. Other simulation parameters are in Table 1.

\begin{figure}[H]
\centering 
\includegraphics[width= 0.85\columnwidth]{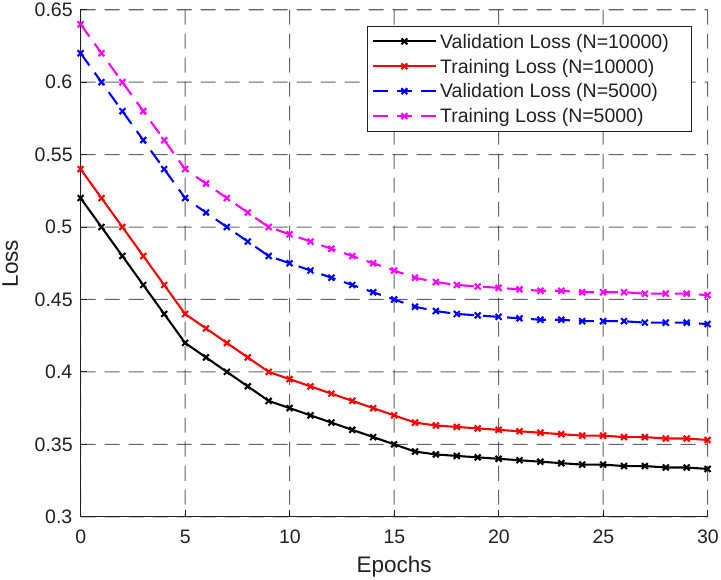}
\caption{ Training and validation for two different dataset sizes. }
\label{ANN loss} 
\end{figure}

The training and validation loss curves in Fig. \ref{ANN loss} indicate that the CNN is effectively learning from the data. For both dataset sizes, the training loss decreases steadily over epochs, demonstrating that the performance of the model improves in the training data. Similarly, the validation loss decreases with the time, indicating that the model generalizes well to the validation data. Furthermore, the CNN model for both dataset sizes avoids over-fitting and delivers acceptable solutions validating its ability to provide solutions in dynamic real-time scenarios. Also, our results show the positive impact of using a larger dataset, which was effectively used.

\begin{figure}[H]
\centering 
\includegraphics[width= 0.73\columnwidth]{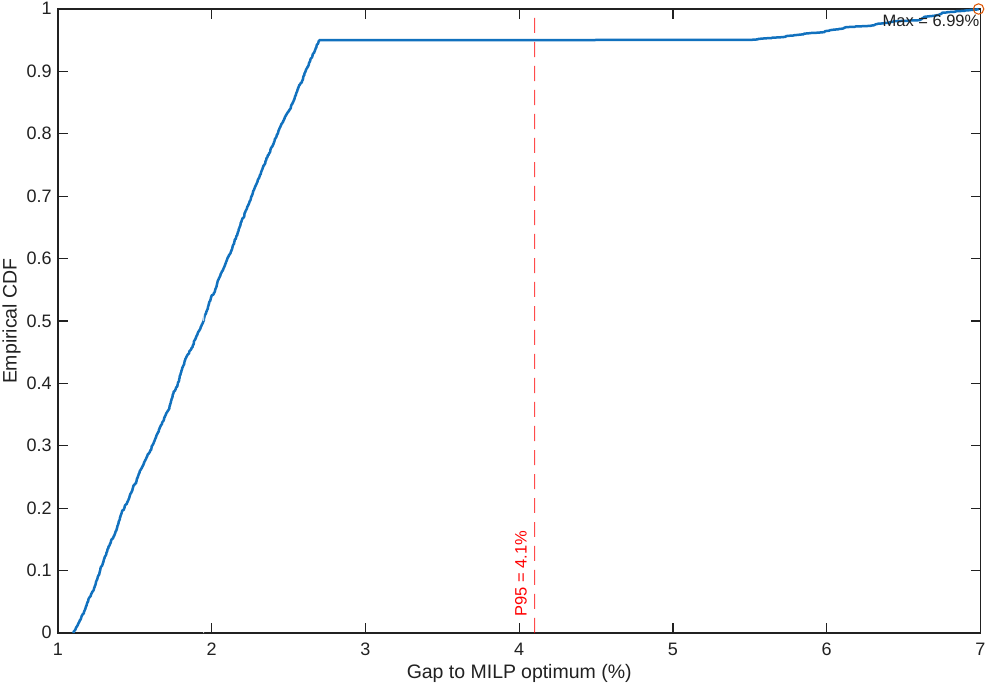}
\caption{CNN-MILP worst case gap}
\label{CNN vs MILP Worst case gap} 
\end{figure}

Interestingly, as shown in Fig. \ref{CNN vs MILP Worst case gap}, the CNN achieves a median gap of $1.9\%$ to the MILP optimum and an empirical worst-case gap no larger than $7\%$ across all mobility, SNR, and load scenarios we evaluated. While deriving a formal upper bound on the performance is challenging given the non-convex nature of both the learning model and the underlying combinatorial problem, these empirical worst-case measurements provide a practical guarantee for deployment and align with our latency/throughput targets. The reported median and worst-case gaps were computed over 1000 held-out test samples covering random user positions, mobility speeds of 0.5–2 m/s, SNRs of 10–30 dB, and user loads of 6–12 users.

\begin{figure}[H]
\centering 
\includegraphics[width= 0.73\columnwidth]{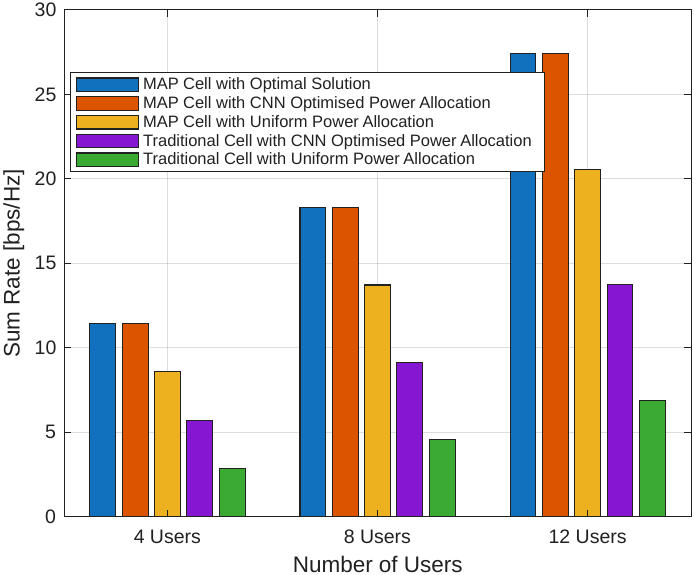}
\caption{Sum rates of different cell schemes. $\gamma = 20dB$, $N = 10 000$.}
\label{ANN Sum Rate} 
\end{figure}

\begin{figure}[H]
\centering 
\includegraphics[width= 0.73\columnwidth]{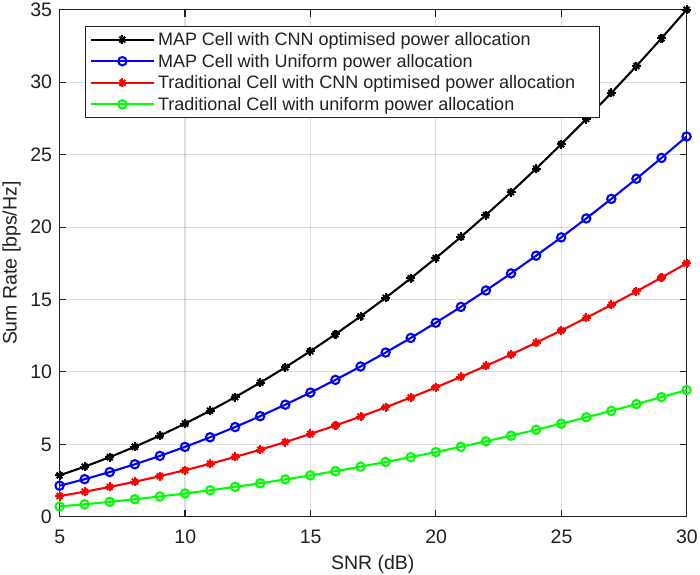}
\vspace{-10pt}
\caption{Sum rates vs SNR. $U = 8$, $N = 10 000$.}
\label{ANN Sum rate vs power} 
\end{figure}

Without loss of generality, we consider a 4-AP MAP cell as shown in Figs. \ref{ANN Sum Rate} and \ref{ANN Sum rate vs power}, other cell sizes can be considered. Our results show that the MAP cell configuration outperforms the traditional scheme in terms of sum rate. This is due to the enhanced signal coverage and power provided by the additional APs operating together. However, ZF may become less effective for managing transmissions as the number of users belonging to a cell increases. The results also show that the CNN-optimised power allocation scheme outperforms the uniform power allocation scheme at different SNRs and numbers of users. The CNN-optimised scheme dynamically adapts to any updates in the network, unlike the uniform approach, which allocates the same power to all users regardless of their conditions. Furthermore, Fig. \ref{ANN Sum Rate} shows that the performance of the overall proposed hybrid ML-based algorithm matches the optimal solution in \cite{10621108}, confirming that the user association and CNN-driven power allocation algorithms work effectively together, enabling decision-making close to the optimal benchmark.

\begin{figure}[H]
\centering 
\includegraphics[width= 0.73\columnwidth]{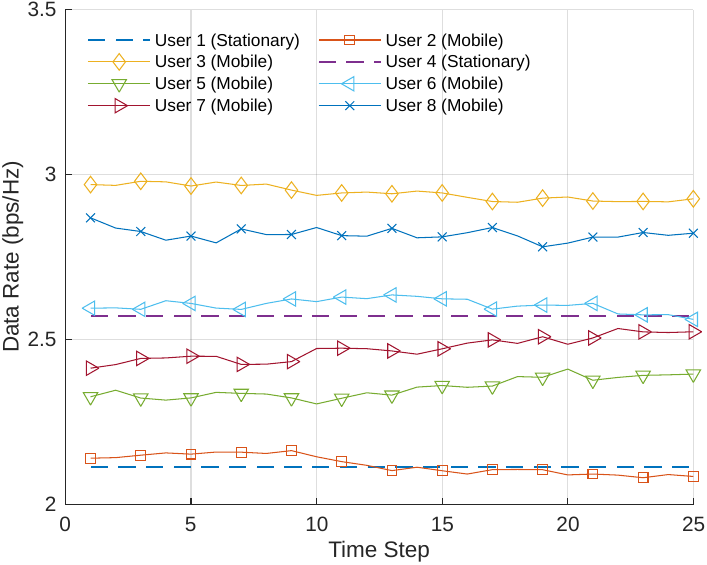}
\caption{User rate versus serving time. Six mobile and two stationary users.}
\label{User mobility} 
\end{figure}

Fig. \ref{User mobility} shows the data rate trends of eight mobile users and reflects how the overall proposed algorithm responds to user mobility in a stable and adaptive manner. It can be seen that users 1 and 4 maintain constant data rates as they remain connected to their respective optical cells at all times. For the remaining users that are mobile, the smooth and gradual fluctuations indicate that the proposed algorithm is able to adapt to changing user positions without causing abrupt shifts in performance. This suggests that the integration of the proposed algorithms for user connectivity and power allocation are effective in maintaining efficiency while avoiding instability in  mobility scenarios.  The overall consistency in the data rates also highlights the robustness and generalisation ability of the CNN in managing dynamic network conditions.

We benchmark the proposed CNN against a fully connected DNN (FC\mbox{-}DNN) and a MILP on a solvable instance of $U{=}8$, $C{=}16$, $\gamma{=}20\,\mathrm{dB}$ as shown in Table \ref{tab:benchmark}. The FC-DNN is included as a baseline because they have been widely used in OWC studies as generic function approximators for ML-based optimisation and performance prediction \cite{FC-DNN1, FC-DNN2, FC-DNN3}, and therefore provide a standard reference for comparison. The MILP provides the reference optimum but requires $\approx 2.3$\,s per instance on a desktop CPU with a GPU, rendering it unsuitable for online use. The FC\mbox{-}DNN (three dense layers, $\sim\!1.20$\,M parameters) attains $96.4\%$ of the MILP sum\mbox{-}rate with median/P95 inference latencies of $0.75/1.30$\,ms. By contrast, our CNN (two $3{\times}3$ convolutional layers followed by 128/64\mbox{-}unit dense layers, $\sim\!0.35$\,M parameters) achieves $98.1\%$ of the MILP optimum with total decision times of $1.7/1.95$\,ms (median/P95). These results are consistent with the complexity analysis: the MILP is accurate but non\mbox{-}scalable online, the FC\mbox{-}DNN is faster yet parameter\mbox{-}heavy, and the CNN offers the best balance of accuracy, parameter efficiency, and latency for real\mbox{-}time OWC control.

\begin{table}[t]
\centering
\caption{Benchmark of CNN, FC-DNN, and MILP ($U{=}8$, $C{=}16$, $\gamma{=}20$\,dB)}
\label{tab:benchmark}
\scriptsize
\setlength{\tabcolsep}{3pt}
\begin{tabular}{|l|c|c|c|c|}
\hline
\textbf{Method} & \textbf{Params (M)} & \textbf{Sum-rate (\%)} & \textbf{Inf.\ Lat.\ (ms)} & \textbf{Total (ms)} \\ \hline
MILP (opt.) & --   & 100   & --            & $\approx 2300$ \\ \hline
FC-DNN      & 1.20 & 96.4  & 0.75 / 1.30   & $\approx 1.4$  \\ \hline
CNN (prop.) & 0.35 & 98.1  & 0.35 / 0.60   & 1.7 / 1.95     \\ \hline
\end{tabular}
\vspace{-6pt}
\end{table}

\section*{Conclusion}
In this work, a sum-rate maximisation problem was formulated for laser-based OWC networks under both conventional and MAP cell formations. Due to the NP-hard nature of the joint user association and power allocation problem, a hybrid ML-based solution was adopted, which comprised of a low-complexity distance-threshold association algorithm and a CNN-based power allocator trained on MILP-optimal solutions. The hybrid scheme achieves near-optimal performance with millisecond-level inference latency, providing a practical alternative to computationally complex optimisation methods. The results further demonstrate that MAP cell formation yields higher throughput compared to conventional cells because of the cooperative VCSEL cells. However, it is anticipated that zero forcing will cease to be as effective as cell size increases.
Future work will explore extensions to more realistic conditions, including imperfect CSI, device-orientation uncertainty, and LoS blockage. In addition, alternative machine learning models such as graph neural networks for association and reinforcement learning for long-term policy optimisation present promising directions for enhancing scalability and adaptability in practical OWC deployments.

\bibliographystyle{IEEEtran}
\bibliography{ANN_Driven_Adaptive_Power_Allocation_for_OWC}

\end{document}